\documentclass[twocolumn,showpacs,preprintnumbers,amsmath,amssymb]{revtex4}
\usepackage{graphicx}
\usepackage{dcolumn}
\usepackage{bm}

\begin{document}

\title{Dynamical clustering in oscillator ensembles with time-dependent
interactions}

\author{Dami\'an H. Zanette
\footnote{Permanent address: Centro At\'omico Bariloche and Instituto Balseiro,
8400 Bariloche, R\'{\i}o Negro, Argentina}}
\email{zanette@cab.cnea.gov.ar}  

\author{Alexander S. Mikhailov}
\email{mikhailov@fhi-berlin.mpg.de} \affiliation{Fritz Haber Institut der Max
Planck Gesellschaft, Abteilung Physikalische Chemie, Faradayweg 4-6, 14195
Berlin, Germany}

\date{\today}

\begin{abstract}

We consider an ensemble of coupled oscillators whose individual states, in
addition to the phase, are characterized by an internal variable with
autonomous evolution. The time scale of this evolution is different for each
oscillator, so that the ensemble is inhomogeneous with respect to the internal
variable. Interactions between oscillators depend on this variable and thus
vary with time. We show that as the inhomogeneity of time scales in the
internal evolution grows, the system undergoes a critical transition between  
ordered and incoherent states. This transition is mediated by a regime of
dynamical clustering, where the ensemble recurrently splits into groups
formed by varying subpopulations. 
\end{abstract}


\maketitle

Synchronization phenomena in interacting dynamical systems have attracted a
great deal of attention in the last two decades \cite{bocca1}.  The spontaneous
appearance of coherent evolution as a result of interactions is ubiquitous in a
vast class of natural and artificial systems, and can be successfully
reproduced by relatively simple models. Most studies of this kind of phenomena
have focused on patterns of highly correlated evolution, such as full and phase
synchronization. Though these forms of synchronization are essential to certain
artificial systems---for instance, secure-communication and chaos-control
devices---they are expected to play a more restricted role in
such objects as biological populations, where the complexity of their
collective function requires a delicate balance between behavioral coherence
and diversity \cite{Mikh}. A clear-cut example  is the human
brain, where highly coherent functional patterns are only realized under
pathological states---notably, during epileptic seizures. Neural tissues, as
well as many natural systems ranging from intracellular molecular complexes to
social populations, are normally found in segregated configurations, where the
system splits into groups of elements with distinct functions. In these
clustered states, highly coherent evolution occurs inside each group, while the
collective dynamics of different groups are much less correlated. This is
typically a dynamical phenomenon, since---as a consequence of the evolution of
the internal state of each element---clusters may temporarily disperse to be
reconstituted later, and individual elements or small groups can migrate
between clusters.  

Clustering has been observed in ensembles of coupled dynamical systems with
specific individual dynamics \cite{kcl,zcl}, or subject to sufficiently
complex interaction functions \cite{golomb,okuda}. Disintegration of clusters
and migration of elements is typically observed under the action of noise 
\cite{kcl,noise,kori}. In this Letter, we show that such kind of
dynamical clustering can occur in an ensemble of coupled periodic oscillators,
in the absence of external fluctuations, when the interaction between
oscillators depends on individual internal state variables with autonomous
evolution.  The time scale of this internal evolution differs between
oscillators. A critical transition between ordered and incoherent evolution,
mediated by dynamical clustering, takes place as the inhomogeneity of such time
scales grows.  

Consider $N$ coupled phase oscillators, each of them characterized by a phase
variable $\phi_i(t)$, obeying by the equations 
\begin{equation} \label{evol}
\dot \phi_i = N^{-1} \sum_{j=1}^N J_{ij}(t) \sin (\phi_j-\phi_i) 
\end{equation}
($i=1,\dots ,N$). When the interaction weights $J_{ij}(t)$ are constant,
$J_{ij}(t) \equiv J$ for all $i$ and $j$, these equations reduce to Kuramoto's
model for identical phase oscillators \cite{kura}, which exhibits full
synchronization for any positive value of $J$. Time-independent non-identical
interaction weights have also been considered, disclosing typical features of
disordered systems, such as glassy-like behavior, frustration, and algebraic
relaxation towards equilibrium \cite{d622,d1073}.  Here, we analyze  the case
where the interaction weights depend on the internal
state of each oscillator, which is specified by a variable $\theta_i(t)$ with
autonomous evolution. Specifically, we consider that $\theta_i$ is itself a
phase variable evolving according to $\dot \theta_i = \Omega_i$, where the
frequency $\Omega_i$ is chosen at random for each element, from a fixed
distribution  $g(\Omega)$. The interaction weight is given by
\begin{equation} \label{J}
J_{ij}(t)= J\cos [\theta_j(t)-\theta_i(t)] ,
\end{equation}
where $J>0$ is a constant factor that can be fixed to $J=1$ by rescaling time. 
With this choice for $J_{ij}$, the interaction between the phase $\phi_i$ and
$\phi_j$ is attractive when $\cos(\theta_j-\theta_i)$ is positive, and
repulsive otherwise. The sign of $J_{ij}$ changes in a time scale of order
$|\Omega_j-\Omega_i|^{-1}$, giving rise---at the level of the whole
ensemble---to a complex time-dependent interaction pattern. This evolving
connection network makes possible the appearance of several dynamical regimes,
including clustering.

We start our study of the above model by performing extensive numerical
realizations in ensembles ranging in size from $N=10^2$ to $10^6$. The internal
frequencies $\Omega_i$ are drawn at random from a Gaussian distribution,
$g(\Omega) = \exp[-(\Omega-\Omega_0)^2/2\sigma^2]/\sqrt{2\pi \sigma^2}$. Since
the dynamics is invariant under a constant shift in these frequencies, we fix
$\Omega_0=0$. It turns out that the collective behavior of the ensemble is
sensible to the frequency dispersion $\sigma$. For small $\sigma$ interactions
evolve slowly, over long time scales. Consequently, the phase $\phi_i$ can
adiabatically follow the evolution of the interaction weights. For long times,
the system is found in a state where the phases $\phi_i$ are homogeneously
distributed over the interval $[0,2\pi)$. A strong correlation is observed in
such state between $\phi_i$ and the internal variable $\theta_i$, namely,
$\phi_i \approx \pm \theta_i + \phi_0$, where $\phi_0$ is a constant.
The sign factor of $\theta_i$ and the value of $\phi_0$ are the same for {\it
all} oscillators, and depend on the initial condition; moreover, $\phi_0$ can
slowly change as time elapses. For large $\sigma$, on the other hand, many of
the interaction weights exhibit large changes over very short times. Each
oscillator is thus subject to rapidly fluctuating forces, which again results
in a state where the phases $\phi_i$ are homogeneously distributed over
$[0,2\pi)$. Now, however, there is no correlation between $\phi_i$ and
$\theta_i$.

\begin{figure}[h]
\centering
\resizebox{\columnwidth}{!}{\includegraphics[clip]{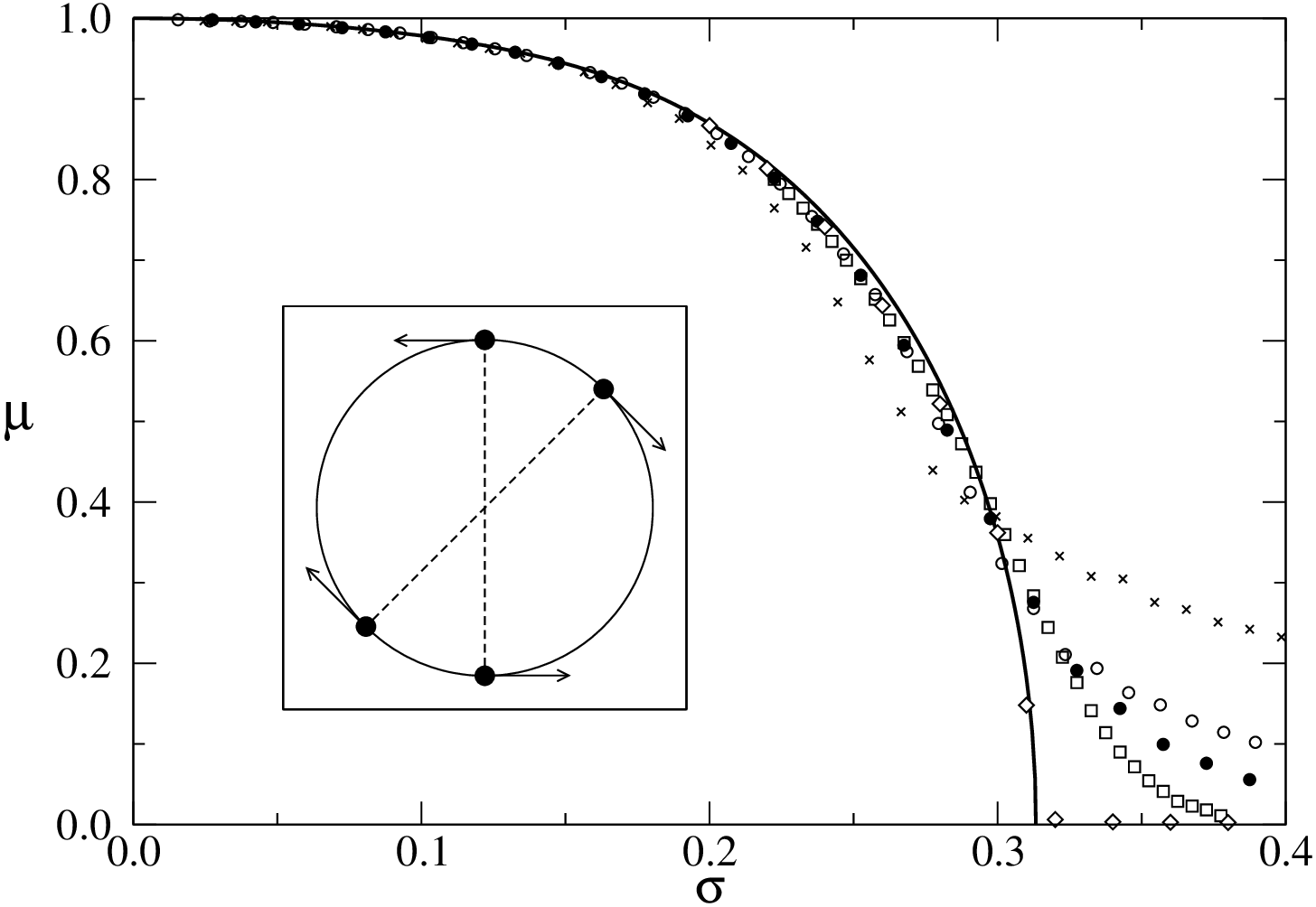}}
\caption{The order parameter $\mu$ as a function of the frequency
dispersion $\sigma$ for ensembles of various sizes, $N=10^2$ $(\times )$,
$10^3$ $(\circ)$, $10^4$ $(\bullet)$, $10^5$ $(\Box)$, and $10^6$ $(\diamond)$.
The curve stands for $r$ as calculated from   Eq.~(\ref{selfc}). Inset:
Schematic representation of the motion of the four clusters observed for
intermediate frequency dispersions, in $\phi$-space (see text).
} \label{f2}
\end{figure}

To characterize the transition between the regimes of small and large frequency
dispersion we introduce, for each oscillator, the two-dimensional complex vector
${\bf m}_i =(\exp[{\rm i}(\phi_i-\theta_i)],\exp[{\rm i}(\phi_i+\theta_i)])$,
and define the average
\begin{equation} \label{bm}
{\bf m} = N^{-1} \sum_{j=1}^N {\bf m}_j \equiv (\mu_+ \exp ({\rm i} \psi_+),
\mu_- \exp ({\rm i} \psi_-)).
\end{equation}
The time average $\mu$ of $\sqrt{\mu_+^2+\mu_-^2}$ is a suitable order parameter
for the transition.  If the system spends most of the time in the state where
$\phi_i\approx \pm \theta_i + \phi_0$ for all $i$, we have $\mu \approx 1$,
while $\mu \sim N^{-1/2}$ in the incoherent state. Figure \ref{f2} shows
numerical results for $\mu$ as a function of $\sigma$ for various system sizes.
The dependence with $N$ suggest the presence of a critical phenomenon for
$\sigma \approx 0.3$.  Within a few assumptions supported by numerical evidence,
this critical behavior can be analytically studied in the limit $N\to \infty$,
as follows. First of all, we note that for $\sigma \neq 0$ and at sufficiently
long times, the internal variables $\theta_i$ are homogeneously distributed
over $[0,2\pi)$. If, in the limit $\sigma\to 0$ (but $\sigma \neq 0$), the
temporal variation of $\theta_i$ is disregarded, $\phi_i = \pm \theta_i
+\phi_0$ is a stationary state of our system, as direct substitution in
Eq.~(\ref{evol}) shows. This conclusion holds for any distribution of
frequencies highly concentrated around $\Omega=0$. For larger values of
$\sigma$, we write $\phi_i = \pm (\theta_i-\delta_i)+\phi_0$, where
$\delta_i(t)$ measures the deviation of each oscillator with respect to the
small-dispersion state. This deviation satisfies
\begin{equation} \label{delta}
\dot \delta_i =  \Omega_i- \frac{\mu}{2} \sin \delta_i,
\end{equation}
where we have written the order parameter $\mu$ in terms of the distribution of
the deviations $\delta_i$ over the ensemble, $p(\delta)$, as
\begin{equation} \label{a}
\mu =  \int_{-\pi/2}^{\pi/2} p(\delta) \cos \delta \ d\delta.
\end{equation}
Here, we have assumed that $p(\delta)$ is an even function; as shown below, this
amounts to postulate that $g(\Omega)$ is itself even.

Equation (\ref{delta}) has been extensively discussed in connection with
phase-oscillator ensembles \cite{saka}. For $|\Omega_i|< \mu/2$ it has a stable
fixed point at one of the solutions of $\sin \delta_i = 2\Omega_i/\mu$. For
$|\Omega_i|>\mu/2$, on the other hand, there are no fixed points and
$|\delta_i|$ grows indefinitely with time. The ensemble can therefore be
thought of as consisting of two subpopulations. Those oscillators for which
$|\Omega_i|<\mu/2$ attain, at asymptotically large times, a stationary
deviation $\delta_i$, which depends on the value of $\Omega_i$. For the
oscillators with $|\Omega_i|>\mu/2$, in contrast, the phase $\phi_i$ does not
reach a stationary value with respect to $\theta_i$. We refer to these two
subpopulations as subensembles I and II, respectively.  The distribution
$p(\delta)$ can be split into contributions from each subensemble, $p=p_{\rm
I}+ p_{\rm II}$. The first contribution is found from the distribution of
frequencies, taking into account the identity $p_{\rm I}(\delta) d\delta=
g(\Omega) d\Omega$. As for $p_{\rm II}$, it can be shown that its contribution
to the integral of Eq.~(\ref{a}) vanishes. This equation becomes, then, 
\begin{equation} \label{selfc}
\mu= \int_{-\mu/2}^{\mu/2} g(\Omega) \sqrt{1-4\Omega^2/\mu^2} \ d\Omega,
\end{equation}
making it possible to find $\mu$ self-consistently  for any (even) distribution
of frequencies. Under quite general conditions on $g(\Omega)$, this equation
describes a second-order transition between disordered ($\mu = 0$) and ordered
($\mu\neq 0$) states. For $\mu\neq 0$, two states are possible,  either with
$\mu_+\neq 0$ or with $\mu_-\neq 0$, which demonstrates the occurrence of
symmetry breaking at the transition. The curve in Fig.~\ref{f2} shows the
solution to Eq.~(\ref{selfc}) for the Gaussian distribution of frequencies
considered in our numerical realizations. It vanishes as $\mu \sim
|\sigma-\sigma_c|^{1/2}$ at $\sigma_c =  \sqrt{\pi/32} \approx 0.313$. 
 
Note that Eq.~(\ref{evol}) can be written, in terms of the average quantity
introduced in Eq.~(\ref{bm}), as
\begin{equation} \label{mf}
\dot \phi_i=\frac{\mu_+}{2} \sin (\psi_++\theta_i-\phi_i)+
\frac{\mu_-}{2} \sin (\psi_- -\theta_i-\phi_i).
\end{equation}
This form of the equation of motion for $\phi_i$ emphasizes the mean-field
nature of interactions. The fact that, as $N\to \infty$, $\mu$ vanishes for
$\sigma>\sigma_c$, implies that the two terms in the right-hand side of
Eq.~(\ref{mf}) vanish as well. For finite system sizes, they are of order
$N^{-1/2}$. Hence, beyond the critical dispersion $\sigma_c$, the evolution of
phases can be thought of as driven by fluctuating forces of order $N^{-1/2}$.
In the thermodynamical limit, for $\sigma > \sigma_c$, the dynamics is frozen.
Below the transition, evolution becomes progressively slower as the critical
point is approached.

\begin{figure}[h]
\centering
\resizebox{\columnwidth}{!}{\includegraphics[clip]{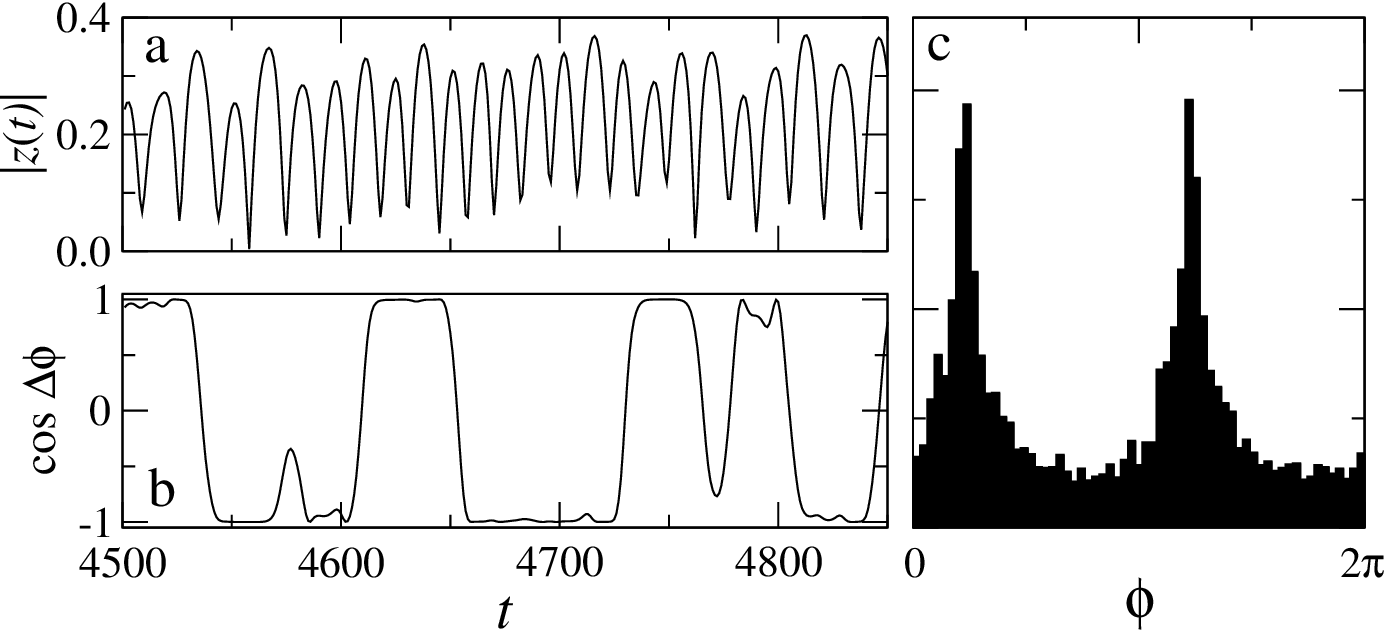}}
\caption{(a) The quantity $|z(t)|$  as a function of time in a system of
$10^4$ oscillators with Gaussian frequency distribution of  dispersion
$\sigma=0.31$. (b) Phase difference of two oscillators with similar frequencies,
in the same realization as in the upper plot. (c) Snapshot of the phase
distribution at one of the maxima of $|z(t)|$, in the same realization.
} \label{f0}
\end{figure}

Numerical simulations show that for intermediate values of $\sigma$, as the
predominance of the ordered small-dispersion state decreases and the critical
transition is approached, a complex regime of clustering develops. Those
oscillators in subensemble II with frequencies just above their lowest value, 
$|\Omega_i|\gtrsim \mu/2$, become divided into four clusters in $\phi$-space.
Though the structure of each cluster is not sharply localized, it is still
possible to define its phase by averaging $\phi_i$ over the oscillators in that
cluster. Two of the clusters are formed by oscillators with $\Omega_i>0$, and
exhibit anti-phase synchronization, so that their phases differ by $\pi$ at all
times. They move monotonically around $\phi$-space. The other two move in the
opposite direction. They are formed by oscillators with $\Omega_i<0$, and also
show anti-phase synchronization (see inset of Fig.~\ref{f2}). Due to the
opposite motion of the two pairs of clusters, they recurrently cross each
other. At these crossings,  their populations are temporarily superimposed in
phase and, at the same time, their motion decelerates and the clusters become
more compact. As a result, two well-localized and relatively long-lived big
clusters, their phases differing by $\pi$,  are recurrently built up out of the
ensemble (see Fig.~\ref{f0}c).

The occurrence of two-cluster states is quantitatively disclosed by the complex
quantity
\begin{equation}
z = N^{-1} \sum_{j=1}^N\exp ( 2{\rm i} \phi_j).
\end{equation}
If the whole ensemble splits into two anti-phase well-localized clusters, we
have $|z(t)| \approx 1$, while for incoherent or higher-order clustered states
we have $|z(t)| \sim N^{-1/2}$. Figure \ref{f0}a shows the evolution of
$|z(t)|$ in the clustering regime, displaying its irregular oscillations
between relatively small and large values. For the same realization,
Fig.~\ref{f0}b displays the phase difference $\Delta \phi$ between two
oscillators in subensemble II, both with positive values of $\Omega_i$, as a
function of time. When $|z(t)|$ is large, $\cos \Delta \phi \approx 1$
identifies a two-cluster state where the two oscillators belong to the same
cluster, while $\cos \Delta \phi \approx -1$ corresponds to the situation where
the oscillators are in different clusters. The intermittent transitions of
$\cos \Delta \phi$ between large positive and negative values reveal the
dynamical nature of clustering, where any two oscillators may alternatively
belong to the same or to different clusters.

\begin{figure}[h]
\centering
\resizebox{\columnwidth}{!}{\includegraphics[clip]{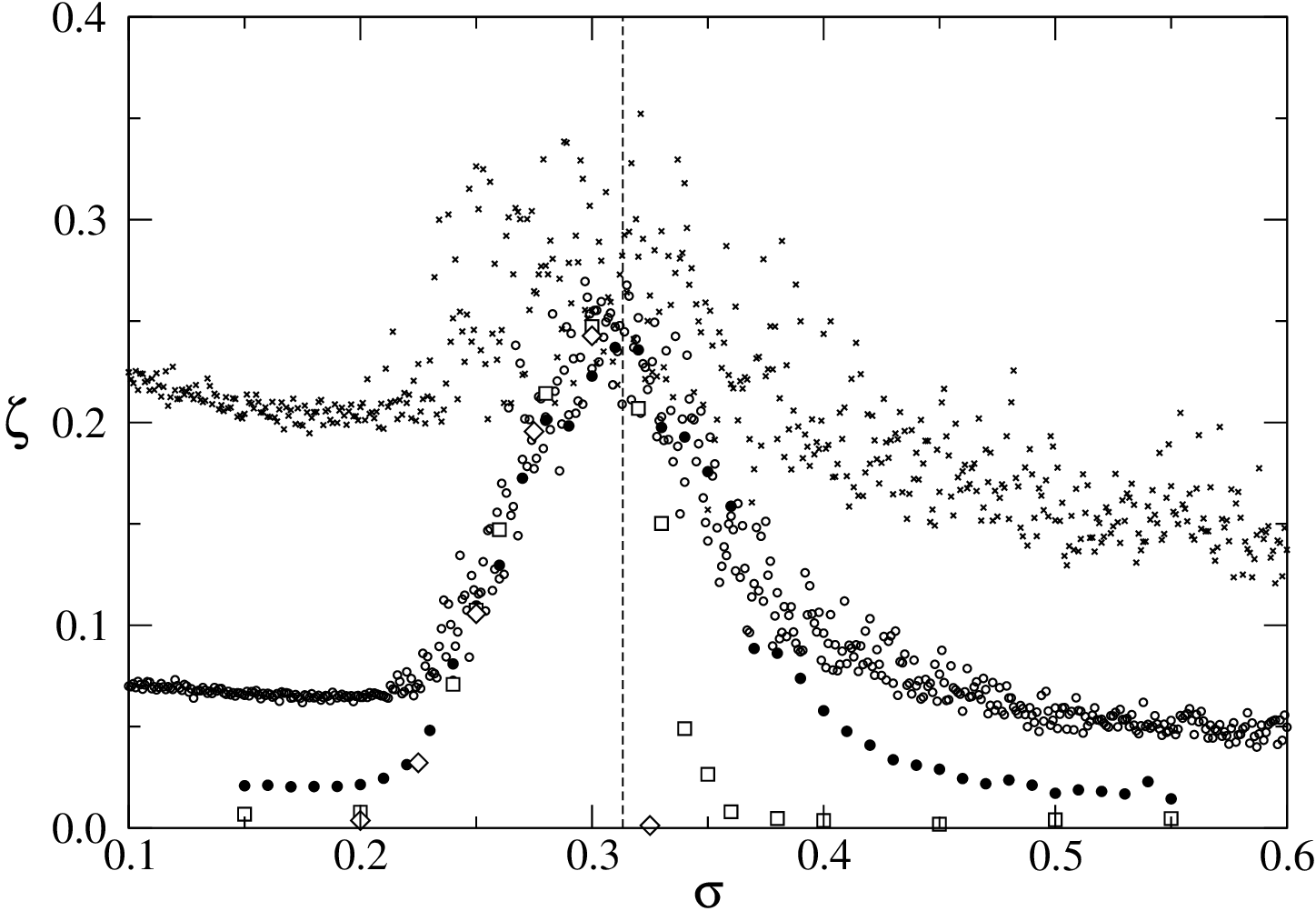}}
\caption{The order parameter $\zeta$ as a function of the frequency dispersion
$\sigma$, obtained from numerical simulations of  ensembles of various sizes,
$N=10^2$ $(\times)$, $10^3$ $(\circ)$,   $10^4$ $(\bullet)$, $10^5$ $(\Box)$,
and $10^6$ $(\diamond)$. The vertical dashed line indicates the critical
dispersion $\sigma_c =\sqrt{\pi/32}$.
} \label{f1}
\end{figure}

The time average $\zeta$ of the quantity $|z(t)|$ plays the role of an order
parameter detecting the two-cluster state.  Figure \ref{f1} displays the
dependence of $\zeta$ with the frequency dispersion $\sigma$, for ensembles of
several sizes. A  peak near $\sigma_c$ becomes clearly defined
as $N$ grows. While for $\sigma>\sigma_c$ the value of $\zeta$ apparently
vanishes for large system sizes, a well-defined profile persists below the
critical point. It grows from zero starting at $\sigma\approx 0.2$ and attains
its maximum, $\zeta \approx 0.25$,  at $\sigma \lesssim \sigma_c$.
  
The formation of the four clusters whose recurrent superimposition gives rise to
the two-cluster states detected by $\zeta$ can be qualitatively understood in
terms of previous results for coupled oscillators with disordered interactions.
Given an oscillator whose internal variable evolves with frequency $\Omega_i$,
the most persistent interactions affecting the dynamics of its phase $\phi_i$
are due to oscillators with similar frequencies, $\Omega_j \approx \Omega_i$.
In this case, in fact, the interaction weights $J_{ij}$ are almost constant
over very long time scales. These weights, however, are still different from
each other, and can take positive and negative values. If the effect of
frustration is moderate, the oscillators become entrained into two anti-phase
groups of similar sizes \cite{d622}. The occurrence of this phenomenon requires
that the oscillator phases are not pinned to the internal variables, and that a
consistent population with similar frequencies $\Omega_i$ actually exists to
trigger condensation. Such conditions are met by those oscillators of
subensemble II with either positive or negative frequencies, close to the limit
$|\Omega_i| = \mu/2$, where the population density is larger. Note that
increasing the frequency dispersion $\sigma$ contributes in two cooperating
ways to the growth  of subensemble II, and thus to the formation of clusters:
it implies, at the same time, a larger population for high frequencies and---as
$\mu$ decreases---a population depletion in subensemble I. As discussed above,
however, clustering is delayed by increasingly long transients as $\sigma$
approaches the critical point. In the limit $N\to \infty$, the evolution
freezes just beyond the critical dispersion, and cluster formation is thus
suppressed.
 
It is possible to give an approximate mathematical description of the motion of
the four clusters, assuming that one of them is centered around a phase
$\phi^*(t)$ and is formed by oscillators whose internal frequencies are, on the
average, $\Omega^*$. The phases and frequencies of the other three clusters can
be derived from symmetry considerations. If the total fraction of the
population involved in clusters is $n^*$, the approximate equation of motion
for the cluster phase reads
\begin{equation}
2\dot \phi^* =-\mu \sin (\phi^*\mp \Omega^*t)-n^*\cos (2\Omega^*t) \sin
(2\phi^*). 
\end{equation}
The two terms in the right-hand side correspond, respectively, to the
contributions of subensembles I and II. They have competing dynamical effects,
and their relative importance changes as $\sigma$ and $n^*$ grow while $\mu$
decreases. The first term favors the development of correlations between
$\phi^*$ and the internal variable $\Omega^* t$.   If, on the other hand, the
second term prevails, the phase difference between clusters of opposite
internal frequencies, $2\phi^*$, spends most of the time close to its two fixed
points, $2\phi^*=0$ or $\pi$, with rapid transitions between them when the sign
of $\cos (2\Omega^*t)$ changes. These fixed points correspond, precisely, to
the mutual crossing of clusters of opposite frequencies that give origin to the
recurrent two-cluster state.

Preliminary numerical results show that dynamical clustering occurs also in
ensembles of particles moving in space and coupled through finite-range
interactions, with autonomous internal variables as those considered above. 
In more complex models of interacting dynamical elements with evolving internal
states, dynamical clustering may be enhanced if the internal variables
themselves undergo segregation into coherent groups. This can be achieved, for
example, if interactions between internal states are allowed. Numerical
realizations suggest that this is the case when the variables $\theta_i$ of the
model studied above are coupled through higher-harmonics interaction functions
\cite{okuda}, which give rise to dynamical clustering if noise is added to the
internal evolution. The same effect has recently been illustrated for ensembles
of interacting particles moving in space, where the internal dynamics of each
element is represented by a chaotic map whose evolution is coupled to those of
neighboring particles \cite{shk}. In real systems---specifically, in biological
populations---the interaction between internal variables is expected to play an
important role when such mechanisms as imitation and social influence are in
action. 

D.~H.~Z.~acknowledges financial support from the Alexander von Humboldt
Stiftung, Germany, and is grateful to the Fritz Haber Institut der Max Planck
Gesellschaft, Berlin, for hospitality.

\end{document}